\documentclass[aps,pra,showpacs,superscriptaddress,preprint]{revtex4}
\usepackage{amssymb}
\usepackage{amsmath}
\usepackage{graphicx}

\catcode`ð=\active
 \defð{\u{g}}
 \catcode`Ð=\active
 \defÐ{\u{G}}
 \catcode`Ý=\active
\defÝ{\. I}
 \catcode`ö=\active
\defö{\"{o}}
 \catcode`Ö=\active
 \defÖ{\"O}
 \catcode`ü=\active
 \defü{\"{u}}
 \catcode`Ü=\active
 \defÜ{\"{U}}
 \catcode`Þ=\active
\defÞ{\c{S}}
 \catcode`þ=\active
 \defþ{\c{s}}
 \catcode`ý=\active
 \defý{{\i}}
 \catcode`ç=\active
\defç{\d{c}}
 \catcode`Ç=\active
\defÇ{\d{C}}

\input{tcilatex}

\begin{document}

\title{A study of quantum pseudodot system with a two-dimensional
pseudoharmonic potential using Nikiforov-Uvarov method}
\author{Sameer M. Ikhdair}
\email[E-mail: ]{sikhdair@neu.edu.tr\\
Tel: +903922236624; Fax: +903922236622}
\date{%
\today%
}

\begin{abstract}
We use the Nikiforov-Uvarov method to calculate the bound states (energy
spectra and wave functions) of a two-dimensional (2D) electron gas
interacted with an exactly solvable pseudoharmonic confinement potential in
a strong uniform magentic field inside dot and Aharonov-Bohm flux field
inside a pseudodot. We give a unified treatment for both Schr\"{o}dinger and
spin-$0$ Klein-Gordon energy spectrum and wave functions as functions of
chemical potential parameter, magnetic field strength, AB flux field and
magnetic quantum number. We obtain analytic expression for the light
interband absorption coefficient and threshold frequency of absorption as
functions of applied magnetic field and geometrical size of quantum
pseudodot. The temperature dependence energy levels for GaAs are also
calculated.

\end{abstract}

\maketitle

\newpage

\section{Introduction}

Over a long time, a considerable interest has been paid for studying size
effects in orbital magnetism [1,2] and the magnetic properties of
low-dimensional metallic and semiconducting structures with restricted
geometries [3] on nanostructures such as dots, wires, wells, antidots, well
wires and antiwells [4,5,6]. Such structures can confine charge carriers in
one, two and three dimensions. Experimental research is currently made to
study the optical and quantum properties of low-dimensional semiconducting
structures for the fabrication purposes and subsequent working of electronic
and optical devices. More studies analyzing these structures have been
focused on the interband light absorption coefficient in the spherical
[7,8,9], parabolic, cylindrical and rectangular [10]\ quantum dots in the
presence and absence of magnetic field [11]. More other works on optical
properties in nanostructures [12,13], band structure calculations, transport
properties of Aharonov-Bohm (AB) type oscillations [14] and
Altshuler-Aharonov-Spivak (AAS) type oscillation [15].

The quantum antidot structure has been modeled in the presence and absence
of repulsive antidot potential, harmonic confining oscillator potential, the
presence and absence of magnetic and Aharonov-Bohm (AB) flux fields in
cylindrical coordinates [16]. This allows one to obtain an exact bound state
solutions for the Schr\"{o}dinger equation. The influence of dots and
antidots on thermodynamic properties (e.g., magnetization) of the system,
the magnetotransport properties and also the magneto-optical (MO)
spectroscopic characteristics of a two-dimensional (2D) electron gas in a
magnetic field are studied in [16]. The nature of MO transitions in this
system demonstrate the appearance of rich spectrum of nonequidistant
frequencies are different from the MO spectrum for a dot modeled by a
harmonic confining potential. The quantum antidot is modeled as an electron
moving outside a cylinder of radius $a$ in the presence of magnetic and AB
flux fields to find analytic expressions for energy and wave function [17].
The numerical and analytical solutions obtained for the dynamics of two
classical electrons interacting via a Coulomb field in a 2D antidot
superlatice potential in the presence of crossed electric and magnetic
fields are quite different than the noninteracting electrons [18]. Some
authors have studied a 2D theoretical model for the quantum dot in which
electrons were confined by a nonhomogenous magnetic field (the so-called
magnetic antidot) [19]. The pseudoharmonic (PH) potential [20,21] is used in
modeling the quantum dots (QDs) and quantum antidots (QADs) in
nanostructures [22]. The spectral properties in a 2D electron confined by a
pseudoharmonic quantum dot (PHQD) potential in the presence of external
strong uniform magnetic field $\overrightarrow{B}$ along the $z$ direction
in the presence of AB flux field created by a selenoid inserted inside the
pseudodot have been studied. The Schr\"{o}dinger and spinless Klein-Gordon
equation are solved exactly for their bound states (energy spectrum and wave
function) [22]. The advantage of the Klein-Gordon solution is that it
provides us relativistic corrections to the commonly known nonrelativistic
solution.

It is well-known that factors such as impurity, electric and magnetic
fields, pressure, and temperature play important roles in the electronic,
optical and transport properties of low-dimensional semiconductor
nanostructures [4,23-28]. In this regard, we carry out detailed exact
analytic analysis of one-particle energetic spectrum and wave functions of
both Schr\"{o}dinger and Klein-Gordon equations with a pseudoharmonic
potential in the presence of magnetic field and Aharonov-Bohm flux field by
using the Nikiforov-Uvarov method [29,30]. The resulting energy spectrum
serves as a base for calculating the corresponding interband light (optical)
absorption coefficient and the threshold frequency value of absorption for
the given model. In addition, the effect of the temperature on the effective
mass is also calculated.

The structure of the paper is as follows. In Sec. 2, the basic formulas of
the Nikiforov-Uvarov (NU) method are outlined in short. In Sec. 3, we
studied the nonrelativistic quantum dot and antidot with the pseudoharmonic
potential in the presence of magnetic and Aharonov-Bohm flux fields. The
exact analytic expressions for the energy spectra and wave functions are
calculated. In Sec. 4, the analytic expressions for the bound states of the
KG electron interacted via the pseudoharmonic potential in the presence of
magnetic field and AB flux field are calculated. These basic formulas are
also reduced to Schr\"{o}dinger solutions for the pseudoharmonic potential
model and free-field interactions under the non-relativistic limits. Results
and discussions are performed in Sec. 5. The conclusions and outlook are
presented in Sec. 6.

\section{Nikiforov-Uvarov Method}

This method is usually used in solving a second-order hypergeometric-type
differential equations satisfying special orthogonal functions [29]. In
spherical or cylindrical coordinates, the resulting Schr\"{o}dinger-like
equation with a given potential is reduced to a hypergeometric type equation
through making a convenient change of variables, say, $r\rightarrow s$ and
then solved systematically for its exact or approximate eigensolutions
(energy levels and wave functions). The most convenient equation, we
consider here, takes the standard form [30] 
\begin{equation}
f^{\prime \prime }(s)+\frac{\widetilde{\tau }(s)}{\sigma (s)}f^{\prime }(s)+%
\frac{\widetilde{\sigma }(s)}{\sigma ^{2}(s)}f(s)=0,
\end{equation}%
where $\sigma (s)$ and $\widetilde{\sigma }(s)$ are polynomials at most of
second order, and $\widetilde{\tau }(s)$ is a first-degree polynomial and $%
f(s)$ is a hypergeometric type polynomial.

Next, we try to reduce Eq. (1) to a more comprehensible form by taking $%
f(s)=\phi (s)y(s)$ and choosing an appropriate function $\phi (s)$: 
\begin{equation}
y^{\prime \prime }(s)+\left( 2\frac{\phi ^{\prime }(s)}{\phi (s)}+\frac{%
\widetilde{\tau }(s)}{\sigma (s)}\right) y^{\prime }(s)+\left( \frac{\phi
^{\prime \prime }(s)}{\phi (s)}+\frac{\phi ^{\prime }(s)}{\phi (s)}\frac{%
\widetilde{\tau }(s)}{\sigma (s)}+\frac{\widetilde{\sigma }(s)}{\sigma
^{2}(s)}\right) y(s)=0.
\end{equation}%
which appears to be more complicated than the standard form given in (1). To
simplify (2), at first, we take the coefficient of $y^{\prime }(s),$ 
\begin{equation}
2\frac{\phi ^{\prime }(s)}{\phi (s)}+\frac{\widetilde{\tau }(s)}{\sigma (s)}=%
\frac{\tau (s)}{\sigma (s)},
\end{equation}%
and set 
\begin{equation}
\frac{\phi ^{\prime }(s)}{\phi (s)}=\frac{\pi (s)}{\sigma (s)},
\end{equation}%
to obtain 
\begin{equation}
\pi (s)=\frac{1}{2}[\tau (s)-\widetilde{\tau }(s)],
\end{equation}%
where $\pi (s)$ is a polynomial of degree at most one. Overmore, the above
equation can be rewritten in the form: 
\begin{equation}
\tau (s)=\widetilde{\tau }(s)+2\pi (s),
\end{equation}%
in which $\tau (s)$ is a polynomial of order one. On the other hand, we can
express the term $\phi ^{\prime \prime }(s)/\phi (s)$ appearing as one of
the coefficients of Eq. (2) as 
\begin{equation}
\frac{\phi ^{\prime \prime }(s)}{\phi (s)}=\left( \frac{\phi ^{\prime }(s)}{%
\phi (s)}\right) ^{\prime }+\left( \frac{\phi ^{\prime }(s)}{\phi (s)}%
\right) ^{2}=\left( \frac{\pi (s)}{\sigma (s)}\right) ^{\prime }+\left( 
\frac{\pi (s)}{\sigma (s)}\right) ^{2}.
\end{equation}%
In this case, the coefficient of $y(s)$ can be simply recasted in the form: 
\begin{equation}
\frac{\phi ^{\prime \prime }(s)}{\phi (s)}+\frac{\phi ^{\prime }(s)}{\phi (s)%
}\frac{\widetilde{\tau }(s)}{\sigma (s)}+\frac{\widetilde{\sigma }(s)}{%
\sigma ^{2}(s)}=\frac{\bar{\sigma}(s)}{\sigma ^{2}(s)}
\end{equation}%
where 
\begin{equation}
\bar{\sigma}(s)=\widetilde{\sigma }(s)+\pi ^{2}(s)+\pi (s)[\widetilde{\tau }%
(s)-\sigma ^{\prime }(s)]+\pi ^{\prime }(s)\sigma (s).
\end{equation}%
Substituting the right-hand sides of Eq. (3) and Eq. (8) into Eq. (2), we
finally obtain 
\begin{equation}
y^{\prime \prime }(s)+\frac{\tau (s)}{\sigma (s)}y^{\prime }(s)+\frac{\bar{%
\sigma}(s)}{\sigma ^{2}(s)}y(s)=0.
\end{equation}%
The above transformation allows one to set the hypergeometric function $%
f(s)=\phi (s)y(s)$, where $\phi (s)$ needs to satisfy the relation (4) with
an arbitrary linear polynomial $\pi (s)$. Thus, making the substitution:%
\begin{equation*}
\bar{\sigma}(s)=\lambdabar \sigma (s),
\end{equation*}%
where $\lambdabar $ is a constant. Hence, Eq. (10) turns into the so-called
hypergeometric type equation: 
\begin{equation}
\sigma (s)y^{\prime \prime }+\tau (s)y^{\prime }+\lambdabar y=0,
\end{equation}%
whose solution is already been given in [31]. Now, comparing Eq. (9) with
Eq. (11) leads to the following quadratic equation: 
\begin{equation}
\pi ^{2}(s)+[\widetilde{\tau }(s)-\sigma ^{\prime }(s)]\pi (s)+\widetilde{%
\sigma }(s)-k\sigma (s)=0,
\end{equation}%
where 
\begin{equation}
k=\lambdabar -\pi ^{\prime }(s).
\end{equation}%
Thus, the solution of quadratic equation (12) is given by 
\begin{equation}
\pi (s)=\frac{\sigma ^{\prime }(s)-\widetilde{\tau }(s)}{2}\pm \sqrt{\left( 
\frac{\sigma ^{\prime }(s)-\widetilde{\tau }(s)}{2}\right) ^{2}-\widetilde{%
\sigma }(s)+k{\sigma }(s)},
\end{equation}%
where the parameter $k$ inside the square root sign must be found explicitly
to enable one to find the physical solutions of Eq. (14) for the plus and
minus signs. Therefore, the expression under the square root sign has to be
the square of a polynomial, since $\pi (s)$ is a polynomial of degree at
most one which provides an equation of the quadratic form available for the
constant $k$. Having set the discriminant of this quadratic equal to zero,
the constant $k$ is determined clearly. Once the constant $k$ is found, the
task of the determination of the polynomial $\pi (s)$ from (14) becomes
simple and straightforward. Further, $\tau (s)$ and $\lambdabar $ can also
be found from Eq. (6) and Eq. (13), respectively.

To make the solutions of Eq. (11) more general, we try to show that all the
derivatives of hypergeometric type function are also of hypergeometric type.
This can be easily acheived by differentiating Eq. (11) and letting $%
v_{1}(s)=y^{\prime }(s)$ 
\begin{equation}
\sigma (s)v_{1}^{\prime {\prime }}(s)+\tau _{1}(s)v_{1}^{\prime }(s)+\mu
_{1}v_{1}(s)=0,
\end{equation}%
where $\tau _{1}(s)=\tau (s)+\sigma ^{\prime }(s)$ and $\mu _{1}=\lambdabar
+\tau ^{\prime }(s)$. $\tau _{1}(s)$ is a polynomial of degree at most one
and $\mu _{1}$ is independent of the variable $s$. Equation (15) is
obviously a hypergeometric type equation again. Further, taking $%
v_{2}(s)=y^{\prime \prime }(s)$ as a new representation and making the
differentiation for the second time, we obtain 
\begin{equation}
\sigma (s)v_{2}^{\prime {\prime }}(s)+\tau _{2}(s)v_{2}^{\prime }(s)+\mu
_{2}v_{2}(s)=0,
\end{equation}%
where 
\begin{equation}
\tau _{2}(s)=\tau _{1}(s)+\sigma ^{\prime }(s)=\tau (s)+2\sigma ^{\prime
}(s),
\end{equation}%
\begin{equation}
\mu _{2}=\mu _{1}+\tau _{1}^{\prime }(s)=\lambdabar +2\tau ^{\prime
}(s)+\sigma ^{\prime \prime }(s).
\end{equation}%
Repeating this process, a general equation of hypergeometric type for $%
v_{n}(s)=y^{(n)}(s)$ is constructed as a family of particular solutions
corresponding to a given $\lambdabar $; 
\begin{equation}
\sigma (s)v_{n}^{\prime {\prime }}(s)+\tau _{n}(s)v_{n}^{\prime }(s)+\mu
_{n}v_{n}(s)=0,
\end{equation}%
and hence the general recurrence relations for $\tau _{n}(s)$ and $\mu _{n}$
can be found as 
\begin{equation}
\tau _{n}(s)=\tau (s)+n\sigma ^{\prime }(s),
\end{equation}%
\begin{equation}
\mu _{n}=\lambdabar +n\tau ^{\prime }(s)+\frac{n(n-1)}{2}\sigma ^{\prime
\prime }(s),
\end{equation}%
respectively. When we set $\mu _{n}=0$, then Eq. (21) becomes 
\begin{equation}
\lambdabar =\lambdabar _{n_{r}}=-n\tau ^{\prime }(s)-\frac{n(n-1)}{2}\sigma
^{\prime \prime }(s),\quad n=0,1,2,\ldots
\end{equation}%
and hence Eq. (19) has a particular solution%
\begin{equation}
y(s)=y_{n}(s)=\frac{B_{n}}{\rho (s)}\frac{d^{n}}{dr^{n}}\left[ \sigma
^{n}(s)\rho (s)\right] ,
\end{equation}%
which is known as the Rodrigues relation of degree $n$ and $\rho (s)$ is the
weight function satisfying 
\begin{equation}
\left[ \sigma (r)\rho (r)\right] ^{\prime }=\tau (r)\rho (r).
\end{equation}%
Finally, to obtain an eigenvalue solution through the NU method, the
relationship between $\lambdabar $ and $\lambdabar _{n_{r}}$ must be set up
by means of Eq. (13) and Eq. (22).

\section{ Nonrelativistic QDs and QADs Influenced by Magnetic and AB Flux
Fields}

\subsection{Exactly solvable bound states}

Consider a two-dimensional ($2D$) single charged electron, $e$ with an
electronic effective mass, $\mu $ interacting via a radially symmetrical dot
(electron) and antidot (hole)$.$ We will study the spectral properties of
such dot and an antidot in a uniform magnetic field, $\overrightarrow{B}=B%
\widehat{z}$ and an AB flux field, applied simultanously. The Schr\"{o}%
dinger equation is given by [32]%
\begin{equation}
\left[ \frac{1}{2\mu }\left( \overrightarrow{p}+\frac{e}{c}\overrightarrow{A}%
\right) ^{2}+V_{\text{conf}}(\vec{r})\right] \psi (\vec{r},\phi )=E\psi (%
\vec{r},\phi ),
\end{equation}%
where $\overrightarrow{A}$ is the vector potential and the repulsive
pseudoharmonic confinement quantum dot (PHQD) potential, $V_{\text{conf}}(%
\vec{r}),$ describing the harmonic quantum dot and antidot structures, $%
V_{D}(r)=V_{0}r^{2}/r_{0}^{2}$ and $V_{AD}(r)=V_{0}r_{0}^{2}/r^{2},$
respectively, is taken as [20,21] 
\begin{equation}
V_{\text{conf}}(\vec{r})=V_{0}\left( \frac{r}{r_{0}}-\frac{r_{0}}{r}\right)
^{2},
\end{equation}%
where $r_{0}$ and $V_{0}$ are the zero point (effective radius) and the
chemical potential. The vector potential $\overrightarrow{A}$ may be
represented as a sum of two terms, $\overrightarrow{A}=\overrightarrow{A}%
_{1}+\overrightarrow{A}_{2}$ such that $\overrightarrow{\nabla }\times 
\overrightarrow{A}_{1}=\overrightarrow{B}$ and $\overrightarrow{\nabla }%
\times \overrightarrow{A}_{2}=0,$ where $\overrightarrow{B}$ $=B\widehat{z}$
is the applied magnetic field, and $\overrightarrow{A}_{2}$ describes the
additional magnetic flux $\Phi _{AB}$ created by a selenoid inserted inside
the antidot (pseudodot). Hence, the vector potentials have azimuthal
components given by [22] 
\begin{equation}
\overrightarrow{A}_{1}=\frac{Br}{2}\widehat{\phi },\text{ }\overrightarrow{A}%
_{2}=\frac{\Phi _{AB}}{2\pi r}\widehat{\phi },\text{ }\overrightarrow{A}%
=\left( \frac{Br}{2}+\frac{\Phi _{AB}}{2\pi r}\right) \widehat{\phi }.
\end{equation}%
Let us consider the 2D cylindrical form of the wave functions: 
\begin{equation}
\psi (\vec{r},\phi )=\frac{1}{\sqrt{2\pi }}e^{im\phi }g(r),\text{ }m=0,\pm
1,\pm 2,\ldots ,
\end{equation}%
where $m$ is the magnetic quantum number. Now, inserting the wave functions
(28) into the Schr\"{o}dinger equation (25), we obtain the following
equation for the radial wave function $g(r)$:%
\begin{equation}
g^{\prime \prime }(r)+\frac{1}{r}g^{\prime }(r)+\left( \nu ^{2}-\frac{\beta
^{2}}{r^{2}}-\gamma ^{2}r^{2}\right) g(r)=0,
\end{equation}%
where we have defined the parameters: 
\begin{subequations}
\begin{equation}
\nu ^{2}=\frac{2\mu }{\hbar ^{2}}\left( E+2V_{0}\right) -\frac{\mu \omega
_{c}}{\hbar }\left( m+\xi \right) ,
\end{equation}%
\begin{equation}
\beta ^{2}=\left( m+\xi \right) ^{2}+a^{2},
\end{equation}%
\begin{equation}
\gamma ^{2}=\frac{2\mu }{\hbar ^{2}}\frac{V_{0}}{r_{0}^{2}}+\left( \frac{\mu
\omega _{c}}{2\hbar }\right) ^{2},
\end{equation}%
where $\xi =\Phi _{AB}/\Phi _{0}$ with the flux quantum $\Phi _{0}=hc/e,$ $%
\omega _{c}=eB/\mu c$ is the cyclotron frequency and $a=k_{F}r_{0}$ with $%
k_{F}=\sqrt{2\mu V_{0}/\hbar ^{2}}$ is the fermi wave vector of the
electron. The magnetic quantum number $m$ relates to the quantum number $%
\beta $ [Eq. (30b)].\tablenotemark[1]%
\tablenotetext[1]{For this system, only
two independent integer quantum numbers are required.} Consequently, the
radial wave function $g(r)$ is required to satisfy the boundary conditions,%
\textit{\ i.e.},$\ $\ $g(0)=0$ and $g($ $r\rightarrow \infty )=0.$ In order
to solve Eq. (29) by NU method, it is necessary to introduce the following
variable $s=r^{2},$ $r\in (0,\infty )\rightarrow $s$\in (0,\infty )$ which
recasts Eq. (29) in the form of hypergeometric type differential equation
(1) as 
\end{subequations}
\begin{equation}
g^{\prime \prime }(s)+\frac{2}{(2s)}g^{\prime }(s)+\frac{1}{(2s)^{2}}\left(
-\gamma ^{2}s^{2}+\nu ^{2}s-\beta ^{2}\right) g(s)=0,
\end{equation}%
where we set $g(r)\equiv g(s).$ Applying the basic ideas of Ref. [30], by
comparing Eq. (31) with Eq. (1) gives us the following polynomials: 
\begin{equation}
\widetilde{\tau }(s)=2,~~~{\sigma }(s)=2s,~~~\widetilde{\sigma }(s)=-\gamma
^{2}s^{2}+\nu ^{2}s-\beta ^{2}.
\end{equation}%
In the present case, if we substitute the polynomials given by Eq. (32) into
Eq. (14), the following equality for the polynomial $\pi (s)$ can be
obtained 
\begin{equation}
\pi (s)=\pm \frac{1}{2}\sqrt{\gamma ^{2}s^{2}+(2k-\nu ^{2})s+\beta ^{2}}.
\end{equation}%
The expression under the square root of the above equation must be the
square of a polynomial of first degree. This is possible only if its
discriminant is zero and the constant parameter $k$ can be determined from
the condition that the expression under the square root has a double zero.
Hence, $k$ is obtained as $k_{+,-}=\nu ^{2}/2\pm \beta \gamma $. In that
case, it can be written in the four possible forms of $\pi (s)$; 
\begin{equation}
\pi (s)=\left\{ 
\begin{array}{cc}
+\left( \gamma s\pm \beta \right) , & \text{for }k_{+}=\frac{1}{2}\nu
^{2}+\beta \gamma , \\ 
-\left( \gamma s\pm \beta \right) , & \text{for }k_{-}=\frac{1}{2}\nu
^{2}-\beta \gamma .%
\end{array}%
\right.
\end{equation}%
One of the four possible forms of $\pi (s)$ must be chosen to obtain an
energy spectrum formula. Therefore, the most suitable form can be
established by the choice:%
\begin{equation*}
\pi (s)=\beta -\gamma s,
\end{equation*}%
for $k_{-}$. The trick in this selection is to find the negative derivative
of $\tau (s)$ given in Eq. (6). Hence, $\tau (s)$ and $\tau ^{\prime }(s)$
are obtained as 
\begin{equation}
\tau (s)=2\left( 1+\beta \right) -2\gamma s,\text{ }\tau ^{\prime
}(s)=-2\gamma <0~.
\end{equation}%
In this case, a new eigenvalue equation becomes 
\begin{equation}
\lambdabar _{n}=2\gamma n,\text{ }n=0,1,2,\ldots
\end{equation}%
where it is beneficial to invite the quantity $\lambdabar _{n}=-n\tau
^{\prime }(s)-\frac{n(n-1)}{2}\sigma ^{\prime \prime }(s)$ in Eq. (22) with $%
n$ is the radial quantum number. Another eigenvalue equation is obtained
from the equality $\lambdabar =k_{-}+\pi ^{\prime }$ in Eq. (13), 
\begin{equation}
\lambdabar =\frac{\nu ^{2}}{2}-\gamma \left( \beta +1\right) .
\end{equation}%
In order to find an eigenvalue equation, the right-hand sides of Eq. (36)
and Eq. (37) must be compared with each other, i.e., $\lambdabar
_{n}=\lambdabar $. In this case the result obtained will depend on $E_{n,m}$
in the closed form: 
\begin{equation}
\nu ^{2}=2\left( 2n+1+\beta \right) \gamma .
\end{equation}%
Upon the substitution of the terms of right-hand sides of Eqs. (30a)-(30c)
into Eq. (38), we can immediately obtain the following expression for the
energy spectrum formula in the presence of PH potential : 
\begin{equation}
E_{n,m}(\xi ,\beta )=\hbar \Omega \left( n+\frac{\left\vert \beta
\right\vert +1}{2}\right) +\frac{1}{2}\hbar \omega _{c}\left( m+\xi \right)
-2V_{0},\text{ }\Omega =\sqrt{\omega _{c}^{2}+4\omega _{D}^{2}},
\end{equation}%
where $\left\vert \beta \right\vert =\sqrt{\left( m+\xi \right) ^{2}+a^{2}}%
>0 $ is an integer and $\omega _{D}=\sqrt{2V_{0}/\mu r_{0}^{2}}.$ We have
two sets of quantum numbers $(n,m,\beta )$ and $(n^{\prime },m^{\prime
},\beta ^{\prime })$ for dot (electron) and antidot (hole), respectively.
Therefore, expression (39) for the energy levels of the electron (hole) may
be readily used for a study of the thermodynamic properties of quantum
structures with dot and antidot in the presence and absence of magnetic
field.

If we ignore the last $-2V_{0}$ term, the above formula becomes the
Bogachek-Landman [16] energy levels, $E_{n,m}(\xi ,\beta )=\hbar \Omega
\left( n+\frac{\left\vert \beta \right\vert +1}{2}\right) +\frac{1}{2}\hbar
\omega _{c}\left( m+\xi \right) ,$ in the presence of dot and antidot
potential. In the absence of pseudoharmonic quantum dot (PHQD), i.e., $%
V_{0}=0,$ $\Omega \rightarrow \omega _{c},$ then $E_{n,m}(\xi )=\hbar \omega
_{c}\left[ n+\frac{1}{2}(\left\vert m+\xi \right\vert +1)\right] +\frac{1}{2}%
\hbar \omega _{c}\left( m+\xi \right) $ which is the formula in the presence
of magnetic and AB flux fields [16]$.$ If we put $\xi =0,$ i.e., in the
absence of AB flux field, we find $E_{n,m}=\hbar \omega _{c}\left[ n+\frac{1%
}{2}(\left\vert m\right\vert +m+1)\right] $ which is the Landau energy
levels [33]. In the absence of magnetic field ($\omega _{c}=0$) and an AB
flux field ($\xi =0$), we find $E_{n,m}=\left( 4\hbar V_{0}/\mu
r_{0}^{2}\right) \left[ n+\left( \sqrt{m^{2}+2\mu V_{0}r_{0}^{2}/\hbar ^{2}}%
+1\right) /2\right] -2V_{0}.$ When $m=0$, we have $E_{n}=\left( 4\hbar
V_{0}/\mu r_{0}^{2}\right) \left( n+1/2\right) $ for harmonic oscillator
energy spectrum.

Next, we calculate the corresponding wave functions for the present PH
potential model. We find the first part of the wave function through Eq.
(4), i.e.,%
\begin{equation}
\phi _{m}(s)=\exp \left( \int \frac{\pi (s)}{\sigma (s)}ds\right)
=s^{\left\vert \beta \right\vert /2}e^{-\gamma s/2}.
\end{equation}%
Then, the weight function defined by Eq. (24) as 
\begin{equation}
\rho (s)=\frac{1}{\sigma (s)}\exp \left( \int \frac{\tau (s)}{\sigma (s)}%
ds\right) =s^{\left\vert \beta \right\vert }e^{-\gamma s},
\end{equation}%
which gives the second part of the wave function (Rodrigues formula);
namely, Eq.(23): 
\begin{equation}
y_{n,m}(s)\sim s^{-\left\vert \beta \right\vert }e^{\gamma s}\frac{d^{n_{r}}%
}{ds^{n_{r}}}\left( s^{n+\left\vert \beta \right\vert }e^{-\gamma s}\right)
\sim L_{n}^{\left( \left\vert \beta \right\vert \right) }\left( \gamma
s\right) ,
\end{equation}%
where $L_{a}^{\left( b\right) }\left( x\right) =\frac{\left( a+b\right) !}{%
a!b!}F\left( a,b+1;x\right) $ is the associated Laguarre polynomial and $%
Fa,b;x)$ is the confluent hypergeometric function. Using $g(s)=\phi
_{m}(s)y_{n,m}(s),$ in this way we may write the radial wave function in the
following fashion%
\begin{equation}
g(r)=C_{n,m}r^{\left\vert \beta \right\vert }e^{-\gamma r^{2}/2}F\left(
-n,\left\vert \beta \right\vert +1;\gamma r^{2}\right) ,
\end{equation}%
and finally the total wave function (28) becomes%
\begin{equation*}
\psi _{n,m}(\vec{r},\phi )=\sqrt{\frac{\gamma ^{\left\vert \beta \right\vert
+1}n!}{\pi \left( n+\left\vert \beta \right\vert \right) !}}r^{\left\vert
\beta \right\vert }e^{-\gamma r^{2}/2}L_{n}^{\left( \left\vert \beta
\right\vert \right) }\left( \gamma r^{2}\right) e^{im\phi }
\end{equation*}%
\begin{equation}
=\frac{1}{\left\vert \beta \right\vert !}\sqrt{\frac{\gamma ^{\left\vert
\beta \right\vert +1}\left( n+\left\vert \beta \right\vert \right) !}{\pi n!}%
}r^{\left\vert \beta \right\vert }e^{-\gamma r^{2}/2}F\left( -n,\left\vert
\beta \right\vert +1;\gamma r^{2}\right) e^{im\phi }.
\end{equation}%
The energy levels in Eq. (39) differ from the usual Landau levels in
cylindrical coordinate system [34] to which it transforms when $\xi =0$
(i.e., $\Phi _{AB}=0$), and $a\rightarrow 0$ (i.e., when the chemical
potential of dot and antidot vanishes, i.e., $V_{0}\rightarrow 0$).
Nevertheless, the Landau levels are nearly continuous discrete spectrum for
a particle confined to a large box with $B=0$ to equally spaced levels
corresponding to $B>0.$ Each increment of energy, $\hbar \omega _{c},$
corresponding to free particle states, which is the degeneracy of each
Landau level leading to a larger spacing as magnetic field $B$ tends to
become stronger [33]. The present model removes this degeneracy with energy
levels spectrum becomes%
\begin{equation}
E_{n,m}=\hbar \omega _{c}\left[ n+\frac{1}{2}\left( \left\vert m\right\vert
+m+1\right) \right] ,
\end{equation}%
and the wave function reads as%
\begin{equation}
\psi _{n,m}(\vec{r},\phi )=\frac{1}{m!}\sqrt{\frac{\gamma ^{m+1}\left(
n+m\right) !}{\pi n!}}r^{m}e^{-\gamma r^{2}/2}F\left( -n,m+1;\gamma
r^{2}\right) e^{im\phi },
\end{equation}%
where $\gamma =(\mu \omega _{c})/2\hbar .$ In the limit when $\omega _{c}\ll
g=\sqrt{\frac{8V_{0}}{\mu }}\frac{c}{r_{0}},$then we have%
\begin{equation}
E_{nm}=\varepsilon _{0}+\varepsilon _{1}\omega _{c}+\varepsilon _{2}\omega
_{c}^{2}-\varepsilon _{4}\omega _{c}^{4}+...,\text{ }
\end{equation}%
where%
\begin{equation}
\varepsilon _{0}=-2V_{0}+N_{nm}g,\text{ }\varepsilon _{1}=\frac{\hbar m}{2},%
\text{ }\varepsilon _{2}=\frac{N_{nm}}{2g},\text{ }\varepsilon _{4}=\frac{%
N_{nm}}{8g^{3}},\text{ }N_{nm}=\hbar \left( n+\frac{m+1}{2}\right) ,\text{ }%
g=\frac{1}{r_{0}}\sqrt{\frac{8V_{0}}{\mu }}.
\end{equation}

\subsection{Interband light absorption coefficient}

Expressions (39) and (44), obtained above for charge carriers (electron or
hole) energy spectrum and the corresponding wave function in quantum
pseudodot under the influence of external magnetic field and AB flux field,
allow to calculate the direct interband light absorption coefficient $K(%
\overline{\omega })$ in such system and the threshold frequency of
absorption. The light absorption coefficient can be expressed as [11,12,35]: 
\begin{equation*}
K(\overline{\omega })=N\dsum\limits_{n,m,\beta }\dsum\limits_{n^{\prime
},m^{\prime },\beta ^{\prime }}\left\vert \dint \psi _{n,m,\beta }^{e}(\vec{r%
},\phi )\psi _{n^{\prime },m^{\prime },\beta ^{\prime }}^{h}(\vec{r},\phi
)rdrd\phi \right\vert ^{2}\delta \left( \Delta -E_{n,m,\beta
}^{e}-E_{n^{\prime },m^{\prime },\beta ^{\prime }}^{h}\right) ,
\end{equation*}%
\begin{equation*}
=N\dsum\limits_{n,m,\beta }\dsum\limits_{n^{\prime },m^{\prime },\beta
^{\prime }}\frac{\gamma ^{\left\vert \beta \right\vert +\left\vert \beta
^{\prime }\right\vert +2}\left( n+\left\vert \beta \right\vert \right)
!\left( n^{\prime }+\left\vert \beta ^{\prime }\right\vert \right) !}{\pi
^{2}n!n^{\prime }!\left( \left\vert \beta \right\vert !\right) ^{2}\left(
\left\vert \beta ^{\prime }\right\vert !\right) ^{2}}\left\vert
\dint_{0}^{2\pi }e^{i\left( m+m^{\prime }\right) \phi }d\phi
\dint_{0}^{\infty }rdre^{-\left( \gamma +\gamma ^{\prime }\right)
r^{2}/2}r^{\left\vert \beta \right\vert +\left\vert \beta ^{\prime
}\right\vert }\right. \text{ }
\end{equation*}%
\begin{equation}
\times \left. F\left( -n,\left\vert \beta \right\vert +1;\gamma r^{2}\right)
F\left( -n^{\prime },\left\vert \beta ^{\prime }\right\vert +1;\gamma
^{\prime }r^{2}\right) \right\vert ^{2}\delta \left( \Delta -E_{n,m,\beta
}^{e}-E_{n^{\prime },m^{\prime },\beta ^{\prime }}^{h}\right) ,
\end{equation}%
where $\Delta =\hbar \overline{\omega }-\varepsilon _{g},$ $\varepsilon _{g}$
is the width of forbidden energy gap, $\overline{\omega }$ is the frequency
of incident light, $N$ is a quantity proportional to the square of dipole
moment matrix element modulus, $\psi ^{e(h)}$ is the wave function of the
electron (hole) and $E^{e(h)}$ is the corresponding energy of the electron
(hole).

Now, we use the integrals [33]%
\begin{equation}
\dint_{0}^{2\pi }e^{i\left( m+m^{\prime }\right) \phi }d\phi =\left\{ 
\begin{array}{ccc}
2\pi & \text{if} & m=-m^{\prime }, \\ 
0 & \text{if} & m\neq -m^{\prime },%
\end{array}%
\right.
\end{equation}%
and%
\begin{equation*}
\dint_{0}^{\infty }e^{-\kappa x}x^{\lambda -1}F\left( -n,\lambda ;qx\right)
F\left( -n^{\prime },\lambda ;q^{\prime }x\right) dx=\Gamma (\lambda )\kappa
^{n+n^{\prime }-\lambda }\left( \kappa -q\right) ^{-n}\left( \kappa
-q^{\prime }\right) ^{-n^{\prime }}
\end{equation*}%
\begin{equation}
\times 
\begin{array}{c}
_{2}F_{1}%
\end{array}%
\left( n,n^{\prime },\lambda ;\frac{qq^{\prime }}{\left( \kappa -q\right)
\left( \kappa -q^{\prime }\right) }\right) ,
\end{equation}%
where $\Gamma (x)$ is the Euler-Gamma function and $%
\begin{array}{c}
_{2}F_{1}%
\end{array}%
\left( a,b,c;z\right) $ is the hypergeometric function, to calculate the
light absorption coefficient:%
\begin{equation}
K(\overline{\omega })=N\dsum\limits_{n,m,\beta }\dsum\limits_{n^{\prime
},m^{\prime },\beta ^{\prime }}P_{n,n^{\prime }}^{\beta }Q_{n,n^{\prime
}}^{\beta }\delta \left( \Delta -E_{n,m,\beta }^{e}-E_{n^{\prime },m^{\prime
},\beta ^{\prime }}^{h}\right) ,
\end{equation}%
where%
\begin{equation}
P_{n,n^{\prime }}^{\beta }=\frac{1}{\left( \left\vert \beta \right\vert
!\right) ^{4}}\left( \gamma \gamma ^{\prime }\right) ^{\left\vert \beta
\right\vert +1}\left( \frac{\gamma +\gamma ^{\prime }}{\gamma -\gamma
^{\prime }}\right) ^{2\left( n+n^{\prime }\right) }\frac{\left( n+\left\vert
\beta \right\vert \right) !\left( n^{\prime }+\left\vert \beta \right\vert
\right) !}{n!n^{\prime }!},
\end{equation}%
and%
\begin{equation}
Q_{n,n^{\prime }}^{\beta }=\left[ \left\vert \beta \right\vert !\left( \frac{%
2}{\gamma +\gamma ^{\prime }}\right) ^{\left\vert \beta \right\vert +1}%
\begin{array}{c}
_{2}F_{1}%
\end{array}%
\left( n,n^{\prime },\left\vert \beta \right\vert +1;-\frac{4\gamma \gamma
^{\prime }}{\left( \gamma -\gamma ^{\prime }\right) ^{2}}\right) \right]
^{2}.
\end{equation}%
Using Eqs. (39) and (49), we find the threshold frequency value of
absorption as%
\begin{equation*}
\hbar \overline{\omega }=\varepsilon _{g}+\hbar \left( n+\frac{\sqrt{\left(
m+\Phi _{AB}/\Phi _{0}\right) ^{2}+2\mu V_{0}r_{0}^{2}/\hbar ^{2}}+1}{2}%
\right) \sqrt{\left( \frac{qB}{\mu c}\right) ^{2}+\frac{8V_{0}}{\mu r_{0}^{2}%
}}+\frac{q\hbar B}{2\mu c}\left( m+\frac{\Phi _{AB}}{\Phi _{0}}\right)
\end{equation*}%
\begin{equation}
+\hbar \left( n^{\prime }+\frac{\sqrt{\left( m^{\prime }+\Phi _{AB}/\Phi
_{0}\right) ^{2}+2\mu ^{\prime }V_{0}r_{0}^{2}/\hbar ^{2}}+1}{2}\right) 
\sqrt{\left( \frac{qB}{\mu ^{\prime }c}\right) ^{2}+\frac{8V_{0}}{\mu
^{\prime }r_{0}^{2}}}+\frac{q\hbar B}{2\mu ^{\prime }c}\left( m^{\prime }+%
\frac{\Phi _{AB}}{\Phi _{0}}\right) -4V_{0}.
\end{equation}%
When $n=m=0,$ then%
\begin{equation*}
\hbar \overline{\omega }_{00}=\varepsilon _{g}+\frac{\hbar }{2}\left( \sqrt{%
\left( \Phi _{AB}/\Phi _{0}\right) ^{2}+2\mu V_{0}r_{0}^{2}/\hbar ^{2}}%
+1\right) \sqrt{\left( \frac{qB}{\mu c}\right) ^{2}+\frac{8V_{0}}{\mu
r_{0}^{2}}}+\frac{q\hbar B}{2\mu c}\frac{\Phi _{AB}}{\Phi _{0}}
\end{equation*}%
\begin{equation}
+\frac{\hbar }{2}\left( \sqrt{\left( m^{\prime }+\Phi _{AB}/\Phi _{0}\right)
^{2}+2\mu ^{\prime }V_{0}r_{0}^{2}/\hbar ^{2}}+1\right) \sqrt{\left( \frac{qB%
}{\mu ^{\prime }c}\right) ^{2}+\frac{8V_{0}}{\mu ^{\prime }r_{0}^{2}}}+\frac{%
q\hbar B}{2\mu ^{\prime }c}\frac{\Phi _{AB}}{\Phi _{0}}-4V_{0}.
\end{equation}

\subsection{Temperature dependence of the effective mass}

The variation of the effective mass with temperature is determined according
to the expression [28,36,37]%
\begin{equation}
\frac{\mu _{e}}{\mu (T)}=\frac{1}{f(T)}=1+E_{p}^{\Gamma }\left[ \frac{2}{%
E_{g}^{\Gamma }(T)}+\frac{1}{E_{g}^{\Gamma }(T)+\Delta _{0}}\right] ,
\end{equation}%
where $\mu _{e}$ is the electronic mass, $E_{p}^{\Gamma }=7.51$ $eV$ is the
energy related to the momentum matrix element, $\Delta _{0}=0.341$ $eV$ is
the spin-orbit splitting and $E_{g}^{\Gamma }(T)$ is the
temperature-dependence of the energy gap (in $eV$ units) at the $\Gamma $
point which is given by [12,36,38,39]%
\begin{equation}
E_{g}^{\Gamma }(T)=1.519-\frac{\left( 5.405\times 10^{-4}\right) T^{2}}{T+204%
}\text{ }(eV).
\end{equation}%
Table 1 lists the temperature-dependent effective mass to the effective mass
of donor electron, i.e., $\mu (T)/\mu _{e}$ for different values of
temperatures. It is seen from Table 1 that raising the temperature will
decrease the value of $f(T)=\mu (T)/\mu _{e}.$ As a matter of fact, the
decrease in this value means that kinetic energy of the donor electron
decrease and consequently lowering the binding energy. The results are
similar to Ref. [28]. Hence the temperature dependence energy spectrum
formula can be expressed as%
\begin{equation*}
E_{n,m}(B,T)=\frac{\hbar \omega _{c}}{f(T)}\left[ \sqrt{1+4\frac{\omega
_{D}^{2}}{\omega _{c}^{2}}f(T)}\left( n+\frac{\sqrt{\left( m+\xi \right)
^{2}+a^{2}f(T)}+1}{2}\right) +\frac{m+\xi }{2}\right] -2V_{0},
\end{equation*}%
which for GaAs turns to be%
\begin{equation*}
E_{n,m}(B,T)=14.9254\hbar \omega _{c}\left[ \sqrt{1+0.268\frac{\omega
_{D}^{2}}{\omega _{c}^{2}}}\left( n+\frac{\sqrt{\left( m+\xi \right)
^{2}+0.067a^{2}}+1}{2}\right) +\frac{m+\xi }{2}\right] -2V_{0},
\end{equation*}%
where we have used $\mu =0.067\mu _{e}.$

\section{The Spinless Klein-Gordon Particle in Magnetic and AB Flux Fields}

The Klein-Gordon (KG) equation is wave equation mostly used in describing
particle dynamics in relativistic quantum mechanics. Nonetheless, physically
this equation describes a scalar particle (spin $0$). Moreover, this wave
equation, for free particles, is constructed using two objects: the
four-vector linear momentum operator $P_{\mu }=i\hbar \partial _{\mu }$ and
the scalar rest mass $M,$ allows one to introduce naturally two types of
potential coupling. One is the gauge-invariant coupling to the four-vector
potential $\left\{ A_{\mu }\left( \overrightarrow{r}\right) \right\} _{\mu
=0}^{3}$ which is introduced via the minimal substitution $P_{\mu
}\rightarrow P_{\mu }-gA_{\mu },$ where $g$ is a real coupling parameter.
The other, is an additional coupling to the space-time scalar potential $S_{%
\text{conf}}(\overrightarrow{r})$ which is introduced by the substitution $%
M\rightarrow M+S_{\text{conf}}(\overrightarrow{r}).$ The term
\textquotedblleft four-vector\textquotedblright\ and \textquotedblleft
scalar\textquotedblright\ refers to the corresponding unitary irreducible
representation of the Poincar$\mathbf{{\acute{e}}}$ space-time symmetry
group (the group of rotations and translations in ($3+1$)-dimensional
Minkowski space-time). Gauge invariance of the vector coupling allows for
the freedom to fix the gauge (eliminating the non physical gauge modes)
without altering the physical content of the problem. Many choose to
simplify the solution of the problem by taking the space component of the
vector potential to vanish (i.e., $\overrightarrow{A}$). One may write the
time-component of the four-vector potential as $gA_{0}=V_{\text{conf}}(\vec{r%
}),$ then it ends up with two independent potential functions in the KG
equation. These are the \textquotedblleft vector\textquotedblright\
potential $V_{\text{conf}}(\overrightarrow{r})$ and the \textquotedblleft
scalar\textquotedblright\ potential $S_{\text{conf}}(\overrightarrow{r})$
[40,41]$.$

The free KG equation is written as 
\begin{equation}
(\partial ^{\mu }\partial _{\mu }+M^{2})\psi _{KG}(t,\overrightarrow{r})=0.
\end{equation}%
Moreover, the vector and scalar couplings mentioned above introduce
potential interactions by mapping the free KG equation as 
\begin{equation}
\left\{ c^{2}\left( \overrightarrow{p}+\frac{e}{c}\overrightarrow{A}\right)
^{2}+\left[ M+S_{\text{conf}}(\vec{r})\right] ^{2}\right\} \psi (%
\overrightarrow{r},\phi )=\left[ E-V_{\text{conf}}(\vec{r})\right] ^{2}\psi (%
\overrightarrow{r},\phi ),
\end{equation}%
where $\psi (\overrightarrow{r},\phi )$ is $2D$ cylindrical wave function
defined as in (28). This type of coupling attracted a lot of attention in
the literature due to the resulting simplification in the solution of the
relativistic problem. The scalar-like potential coupling is added to the
scalar mass so that in case when $S_{\text{conf}}(\vec{r})=\pm V_{\text{conf}%
}(\overrightarrow{r}),$ the KG equation could always be reduced to a Schr%
\"{o}dinger-type second order differential equation as follows%
\begin{equation}
\left[ c^{2}\left( \overrightarrow{p}+\frac{e}{c}\overrightarrow{A}\right)
^{2}+2\left( E\pm Mc^{2}\right) V_{\text{conf}}(\overrightarrow{r}%
)+M^{2}c^{4}-E^{2}\right] \psi (\overrightarrow{r},\phi )=0.
\end{equation}%
Hence, the bound state solutions of the above two cases are to be treated
separetely as follows.

\subsection{The $S_{\text{conf}}(\protect\overrightarrow{r})=+V_{\text{conf}%
}(\protect\overrightarrow{r})$ case}

The positive energy states (corresponding to $S_{\text{conf}}(%
\overrightarrow{r})=+V_{\text{conf}}(\overrightarrow{r})$ in the
nonrelativistic limit (taking $E-Mc^{2}\cong E$ and $E+Mc^{2}\cong 2\mu
c^{2},$ where $\left\vert E\right\vert \ll Mc^{2}$) are solutions of%
\begin{equation}
\left[ \frac{1}{2\mu }\left( \overrightarrow{p}+\frac{e}{c}\overrightarrow{A}%
\right) ^{2}+2V_{\text{conf}}(\overrightarrow{r})-E\right] \psi (%
\overrightarrow{r},\phi )=0.
\end{equation}%
where $\psi (\overrightarrow{r},\phi )$ stands for either $\psi ^{(+)}(%
\overrightarrow{r},\phi )$ or $\psi ^{(\text{KG})}(\overrightarrow{r},\phi ).
$ This is the Schr\"{o}dinger equation for the potential $2V_{\text{conf}}(%
\overrightarrow{r}).$ Thus, the choice $S_{\text{conf}}(\overrightarrow{r}%
)=+V_{\text{conf}}(\overrightarrow{r})$ produces a nontrivial
nonrelativistic limit with a potential function $2V_{\text{conf}}(%
\overrightarrow{r}),$ and not $V_{\text{conf}}(\overrightarrow{r}).$
Accordingly, it would be natural to scale the potential term in Eq. (61) and
Eq. (62) so that in the non-relativistic limit the interaction potential
becomes $V_{\text{conf}},$ not $2V_{\text{conf}}.$ thus, we need to recast
Eq. (61) and Eq. (62) as [41] 
\begin{subequations}
\begin{equation}
\left[ c^{2}\left( \overrightarrow{p}+\frac{e}{c}\overrightarrow{A}\right)
^{2}+\left( E+Mc^{2}\right) V_{\text{conf}}(\overrightarrow{r}%
)+M^{2}c^{4}-E^{2}\right] \psi (\overrightarrow{r},\phi )=0.
\end{equation}%
\begin{equation}
\left[ \frac{1}{2\mu }\left( \overrightarrow{p}+\frac{e}{c}\overrightarrow{A}%
\right) ^{2}+V_{\text{conf}}(\overrightarrow{r})-E\right] \psi _{nm}(%
\overrightarrow{r},\phi )=0,
\end{equation}%
with $V_{\text{conf}}(\overrightarrow{r})$ and $\overrightarrow{A}$ are
given in Eq. (26) and Eq. (27), respectively. To avoid repeatition in
solving Eq. (63a), we follow the same steps of solution explained before by
taking $\psi _{nm}(\overrightarrow{r},\phi )=g(r)e^{im\phi }/\sqrt{2\pi }$
to obtain an equation satisfying the radial part: of the wave function: 
\end{subequations}
\begin{equation}
g^{\prime \prime }(s)+\frac{2}{\left( 2s\right) }g^{\prime }(s)+\left(
-b_{1}^{2}s^{2}+\lambda _{1}^{2}s-a_{1}^{2}\right) g(s)=0,
\end{equation}%
where we have used 
\begin{subequations}
\begin{equation}
\lambda _{1}^{2}=\frac{1}{\hbar ^{2}c^{2}}\left[ E^{2}+2\left(
E+Mc^{2}\right) V_{0}-M^{2}c^{4}\right] -\frac{M\omega _{c}}{\hbar }\left(
m+\xi \right) ,
\end{equation}%
\begin{equation}
a_{1}^{2}=\left( m+\xi \right) ^{2}+\frac{r_{0}^{2}}{\hbar ^{2}c^{2}}\left(
E+Mc^{2}\right) V_{0},
\end{equation}%
\begin{equation}
b_{1}^{2}=\left( \frac{M\omega _{c}}{2\hbar }\right) ^{2}+\frac{1}{\hbar
^{2}c^{2}r_{0}^{2}}\left( E+Mc^{2}\right) V_{0}.
\end{equation}%
The solution of Eq. (64) can be easily constructed on making the changes: $%
\nu \rightarrow \lambda _{1},$ $\beta \rightarrow a_{1},$ and $\gamma
\rightarrow b_{1}.$ Thus, the equation for the KG positive energy states can
be easily found from Eq. (38) as 
\end{subequations}
\begin{equation}
\lambda _{1}^{2}=2\left( 2n+1+a_{1}\right) b_{1},
\end{equation}%
and further inserting Eqs. (65a)-(65c), we finally obtain the transcendental
energy formula%
\begin{equation*}
\hbar \left( 1+2n+\sqrt{m^{\prime }{}^{2}+\frac{a^{\prime 2}}{2M}\gamma _{1}}%
\right) \sqrt{\omega _{c}^{2}+\text{ }\frac{\text{2}\omega _{D}^{\prime 2}}{M%
}\gamma _{1}}=\frac{1}{M}\left( \gamma _{2}+2V_{0}\right) \gamma _{1}-\hbar
\omega _{c}m^{\prime },
\end{equation*}%
\begin{equation}
\gamma _{1}=\frac{\left( E+Mc^{2}\right) }{c^{2}},\text{ }\gamma
_{2}=E-Mc^{2},\text{ }\omega _{D}^{\prime }=\sqrt{\frac{2V_{0}}{Mr_{0}^{2}}},%
\text{ }a^{\prime 2}=\frac{2MV_{0}r_{0}^{2}}{\hbar ^{2}}
\end{equation}%
where $m^{\prime }=m+\xi $ is a new quantum number$.$ We may find solution
to the above transcendental equation as $E=E_{KG}^{(+)}.$ In the
nonrelativistic limit ($\gamma _{1}\rightarrow 2M$ and $\gamma
_{2}\rightarrow E$), the above equation can be easily reduced to the simple
energy spectrum formula given in Eq. (39). Overmore, under the above
parameters mapping, the $2D$ KG wave function can be found directly from Eq.
(44) as%
\begin{equation}
\psi _{n,m}^{(+)}(\vec{r},\phi )=\sqrt{\frac{b_{1}^{\left\vert
a_{1}\right\vert +1}n!}{\pi \left( n+\left\vert a_{1}\right\vert \right) !}}%
r^{\left\vert a_{1}\right\vert
}e^{-b_{1}r^{2}/2}L_{n}^{(a_{1})}(b_{1}r^{2})e^{im\phi }.
\end{equation}

\subsection{The $S_{\text{conf}}(\protect\overrightarrow{r})=-V_{\text{conf}%
}(\protect\overrightarrow{r})$ case}

In this case, we follow the same steps of solution in the previous
subsection:%
\begin{equation}
g^{\prime \prime }(s)+\frac{2}{\left( 2s\right) }g^{\prime }(s)+\left(
-b_{2}^{2}s^{2}+\lambda _{2}^{2}s-a_{2}^{2}\right) g(s)=0,
\end{equation}%
where we have used 
\begin{subequations}
\begin{equation}
\lambda _{1}^{2}\rightarrow \lambda _{2}^{2}=\frac{1}{\hbar ^{2}c^{2}}\left[
E^{2}+2\left( E-Mc^{2}\right) V_{0}-M^{2}c^{4}\right] -\frac{M\omega _{c}}{%
\hbar }\left( m+\xi \right) ,
\end{equation}%
\begin{equation}
a_{1}^{2}\rightarrow a_{2}^{2}=\left( m+\xi \right) ^{2}+\frac{r_{0}^{2}}{%
\hbar ^{2}c^{2}}\left( E-Mc^{2}\right) V_{0},
\end{equation}%
\begin{equation}
b_{1}^{2}\rightarrow b_{2}^{2}=\left( \frac{M\omega _{c}}{2\hbar }\right)
^{2}+\frac{1}{\hbar ^{2}c^{2}r_{0}^{2}}\left( E-Mc^{2}\right) V_{0}.
\end{equation}%
Thus, the equation for the KG negative energy states can be readily found as 
\end{subequations}
\begin{equation}
\lambda _{2}^{2}=2\left( 2n+1+a_{2}\right) b_{2},
\end{equation}%
which provides the transcendental energy spectrum formula%
\begin{equation*}
\left( \left( 2n+1\right) \hbar c+\sqrt{\hbar ^{2}c^{2}\left( m+\xi \right)
^{2}+r_{0}^{2}V_{0}\gamma _{2}}\right) \sqrt{M^{2}\omega _{c}^{2}+\frac{%
4V_{0}}{r_{0}^{2}}\gamma _{2}}
\end{equation*}%
\begin{equation}
=\left( c^{2}\gamma _{1}+2V_{0}\right) \gamma _{2}-\hbar cM\omega _{c}\left(
m+\xi \right) ,
\end{equation}%
and the corresponding $2D$ KG wave function is found as%
\begin{equation}
\psi _{n,m}^{(-)}(\vec{r},\phi )=\sqrt{\frac{b_{2}^{\left\vert
a_{2}\right\vert +1}n!}{\pi \left( n+\left\vert a_{2}\right\vert \right) !}}%
r^{\left\vert a_{2}\right\vert
}e^{-b_{2}r^{2}/2}L_{n}^{(a_{2})}(b_{2}r^{2})e^{im\phi }.
\end{equation}%
It should be noted that the negative energy states (corresponding to $S_{%
\text{conf}}(\overrightarrow{r})=-V_{\text{conf}}(\overrightarrow{r})$) are
free fields since under these conditions Eq. (61) reduces to%
\begin{equation}
\left[ -\frac{1}{2\mu }\left( \overrightarrow{p}+\frac{e}{c}\overrightarrow{A%
}\right) ^{2}+E\right] \psi _{n,m}(\overrightarrow{r},\phi )=0.
\end{equation}%
which is a simple free-interaction mode. For the free fields, Eq. (74), the
set of parameters in Eqs. (70a)-(70c) reads%
\begin{equation}
\lambda _{2}=\sqrt{\frac{2\mu E}{\hbar ^{2}}-\frac{\mu \omega _{c}}{\hbar }%
\left( m+\xi \right) },\text{ }a_{2}=m+\xi ,\text{ }b_{2}=\frac{\mu \omega
_{c}}{2\hbar },
\end{equation}%
which lead to the energy spectrum formula%
\begin{equation}
E^{(-)}=\left( n+m+\xi +\frac{1}{2}\right) \hbar \omega _{c},
\end{equation}%
and and wave function%
\begin{equation}
\psi _{nm}^{(-)}(\vec{r},\phi )=\sqrt{\frac{\left( \frac{\mu \omega _{c}}{%
2\hbar }\right) ^{m+\xi +1}n!}{\pi \left( n+m+\xi \right) !}}r^{m+\xi }e^{-%
\frac{\mu \omega _{c}}{4\hbar c}r^{2}}L_{n}^{(m+\xi )}\left( \frac{\mu
\omega _{c}}{2\hbar }r^{2}\right) e^{im\phi }.
\end{equation}

\section{Results and Discussion}

We solved exactly the Schr\"{o}dinger and Klein-Gordon equations for an
electron under the pseudoharmonic interaction consisting of quantum dot
potential and antidot potential in the presence of a uniform strong magnetic
field $\overrightarrow{B}$ along the $z$ axis and AB flux field created by
an infinitely long selenoid inserted inside the pseudodot. We obtained bound
state solutions including the energy spectrum formula (39) and wave function
(44) for a Schr\"{o}dinger electron. Overmore, for the Klein-Gordon
electron, the positive energy equation (67) and wave function (68) is found
for $S_{\text{conf}}(\overrightarrow{r})=+V_{\text{conf}}(\overrightarrow{r}%
) $ case. However, the negative energy equation (67) and wave function (68)
are found for $S_{\text{conf}}(\overrightarrow{r})=+V_{\text{conf}}(%
\overrightarrow{r})$ case. These two cases are reduced to the Schr\"{o}%
dinger equation with a potential interaction $V_{\text{conf}}(%
\overrightarrow{r})$ and free field interaction solutions, respectively. Now
we study the effect of the pseudoharmonic potential, the presence and
absence of magnetic field $B,$ the presence and absence of AB flux density $%
\xi $ and the antidot potential on the energy levels (39). To see the
dependence of the energy spectrum on the magnetic quantum number, $m$, we
take the following values: magnetic field $\overrightarrow{B}=\left( 6\text{ 
}T\right) $ $\widehat{z},$ AB flux field $\xi =8$, chemical potential $%
V_{0}=0.68346$ ($meV)$ and $r_{0}=8.958\times 10^{-6}$ $cm$ [22]. Thus, we
obtained $a=\sqrt{2\mu V_{0}r_{0}^{2}/\hbar ^{2}}=11.997702,$ $2\omega _{D}=%
\sqrt{8V_{0}/\mu r_{0}^{2}}=0.3280381$ $\omega _{c}$ and $\hbar \omega
=1.05243\hbar \omega _{c}$ [34]$,$ the dependence of the energy spectrum,
(39) on the $n$ and $m$ is given by%
\begin{equation}
\frac{E_{n,m}}{\hbar \omega _{c}}=1.05243\left( n+\frac{\sqrt{\left(
m+8\right) ^{2}+12^{2}}+1}{2}\right) +\frac{1}{2}\left( m+8\right)
-1.9678584,\text{ for }B=6\text{ }T.
\end{equation}%
where $m=0,\pm 1,\pm 2,\ldots $ and $n=0,1,2,\ldots .$ For the lowest ground
state ($n=0$): $E_{0,m}/\hbar \omega _{c}=1.05243\left( \sqrt{\left(
m+8\right) ^{2}+12^{2}}+1\right) /2+\left( m+8\right) /2-1.9678584,$ for $%
B=6 $ $T.$ Overmore, to show the effect of magnetic field $B$ on the energy
spectrum, we take values for parameters $\xi =8,$ $V_{0}=0.68459$ $meV$ and $%
r_{0}=8.958\times 10^{-6}$ $cm$ [22], where $a=\sqrt{2\mu
V_{0}r_{0}^{2}/\hbar ^{2}}=12.007617$ and $4\omega _{D}^{2}=8V_{0}/\mu
r_{0}^{2}=0.120039\times 10^{24}$ $(rad/s)^{2},$ the dependence of energy
levels (39) on the magnetic field becomes%
\begin{equation*}
E_{n,m}\left( meV\right) =0.1157705\sqrt{B^{2}+3.8803305}\left( n+\frac{%
\sqrt{\left( m+8\right) ^{2}+12^{2}}+1}{2}\right)
\end{equation*}%
\begin{equation}
+0.1157705B\left( \frac{m+8}{2}\right) -1.36918.
\end{equation}%
In Figure 1, we plot the pseudodot energy levels in the absence (presence)
of pseudodot potential (i.e., $V_{0}=0$ $\rightarrow $ $a=0$ $\left(
V_{0}\neq 0\rightarrow a=12\right) )$ and in the absence (presence) of AB
flux field $\Phi _{AB}$ (i.e., $\xi =0$ ($\xi =8)$) as a function of
magnetic quantum number $m$ for $B=6$ $T$. As demonstrated in Figure 1, the
Landau energy states [33] (i.e., $V_{0}=0$ $\rightarrow $ $a=0,$ $\xi =0$
and $\ \xi =8)$ are degenerate states (see, long dashed and dotted solid
curves) for negative values of $m,$ however, the pseudodot potential removes
this degeneracy $($case when $V_{0}\neq 0$ $\rightarrow $ $a=12),$ (see,
solid and dotted dashed curves). In the absence of pseudodot potential ($%
a=0) $ and presence of AB flux field ($\xi =8$), the degeneracy still exists
(long dashed line). It is found that the energy levels of PHQD potential are
approximately equal the Landau energy levels for large absolute $m$ values.
However, they are quite different for small absolute $m$ values ($-12\preceq
m\preceq 13$ when $\xi =0$ and $-20\preceq m\preceq 5$ when $\xi =8$)$.$ It
is also noted that as the quantum number $n$ increases ($n>0),$ the curves
are quite similar to Figure 1 but the energy levels are pushed up toward the
positive energy for all values of $m.$ In Figure 2 to Figure 7, we plot the
magnetic field dependence of the ground state energy $E_{0,m}(\xi ,a)$ (in
units of $meV)$ in the presence and absence of pseudodot potential and AB
flux field for several values of magnetic quantum numbers $m=27,35,1,0,-24$
and $-16$, respectively. It is shown in Figure 2 to Figure 7 that pseudodot
energy increases with increasing magnetic field strength. Further, in the
absence of pseudodot potential, magnetic field in the positive $z$ direction
removes the degeneracy for positive $m$ values. In these Figures, the
behavior of pseudodot energy as function of the magnetic field $B$ is shown
in the presence of pseudodot potential and AB flux field (solid curves), in
the absence of pseudodot potential and presence of AB flux field (dotted
curves) and the absence of pseudodot potential and AB flux field (dashed
curves).

To investigate the dependence of the energy levels on temperature, we take
the values of parameters: $B=6$ $T,$ $\xi =8$, $V_{0}=0.68346$ ($meV)$ and $%
r_{0}=8.958\times 10^{-6}$ $cm$ [22]. Hence, the temperature dependence of
the energy levels (in the units of $\hbar \omega _{c})$ at the $\Gamma $
point are given by 
\begin{equation*}
\frac{E_{n,m}(T)}{\hbar \omega _{c}}=\frac{1}{f(T)}\left[ \sqrt{1+\left(
0.32804\right) ^{2}f(T)}\left( n+\frac{\sqrt{\left( m+8\right) ^{2}+144f(T)}%
+1}{2}\right) +\frac{m+8}{2}\right]
\end{equation*}%
\begin{equation}
-1.9678584,
\end{equation}%
where $f(T)$ is calculated in Table 1 at any temperature value. In GaAs, we
have $f(T)=0.067$ [11]$.$ Taking the special values of parameters $\xi =8,$ $%
V_{0}=0.68459$ $meV$ and $r_{0}=8.958\times 10^{-6}$ $cm$ [22], two
parameters (temperature and magnetic field) dependence of the energy levels
(in units of $meV)$ are calculated as%
\begin{equation*}
E_{n,m}\left( B,T\right) =\frac{1}{f(T)}\left[ 0.1157705\sqrt{%
B^{2}+3.8803305f(T)}\left( n+\frac{\sqrt{\left( m+8\right) ^{2}+144f(T)}+1}{2%
}\right) \right.
\end{equation*}%
\begin{equation}
+\left. 0.1157705B\left( \frac{m+8}{2}\right) \right] -1.36918\text{ (units }%
meV\text{)}.
\end{equation}%
which becomes%
\begin{equation*}
E_{n,m}(B)=14.9254\left[ 0.1157705\sqrt{B^{2}+0.26}\left( n+\frac{\sqrt{%
\left( m+8\right) ^{2}+9.648}+1}{2}\right) \right.
\end{equation*}%
\begin{equation}
+\left. 0.1157705B\left( \frac{m+8}{2}\right) \right] -1.36918\text{ (units }%
meV\text{)}.
\end{equation}%
for GaAs. Figure 8 to Figure 12 show the variation of the pseudodot energy
levels (in $meV$) as function of magnetic field $B$ (in $Tesla$) (82) in the
presence of pseudodot potential and AB flux field (solid curves), in the
absence of pseudodot potential and presence of AB flux field (dotted curves)
and the absence of pseudodot potential and AB flux field (dashed curves).for
various values of radial quantum numbers $n$ and magnetic quantum numbers $m$%
. For GaAs case, we consider the following cases (a) $n=m=0,$ (b) $n=5,$ $%
m=0,$ (c) $n=0,$ $m=5,$ (d) $n=0,$ $m=-5$ and (e) $n=5,$ $m=-5$ in Figures 8
to 12$,$ respectively$.$

\section{Conclusions and Outlook}

In this work, we have obtained bound state energies and wave functions of
the KG particle in the field of pseudoharmonic quantum dot and antidot
structure in the presence of a uniform magnetic and AB flux fields. The
positive (negative) KG energy states corresponding to $S_{\text{conf}}(%
\overrightarrow{r})=+V_{\text{conf}}(\overrightarrow{r})$ ($S_{\text{conf}}(%
\overrightarrow{r})=-V_{\text{conf}}(\overrightarrow{r})$) are studied.
Overmore, the Schr\"{o}dinger bound state solutions are found. Under
nonrelativistic limit, the KG equation with equal mixture of scalar and
vector potentials $S_{\text{conf}}(\overrightarrow{r})=+V_{\text{conf}}(%
\overrightarrow{r})$ and $S_{\text{conf}}(\overrightarrow{r})=-V_{\text{conf}%
}(\overrightarrow{r})$ can be easily reduced into the well-known Schr\"{o}%
dinger equation of a particle with an interaction potential field and a free
field, respectively. Overmore, the nonrelativistic electron and hole energy
spectra and the their corresponding wave functions are used to calculate the
the interband light absorption coefficient and the the threshold frequency
of absorption. Also, the energy spectrum of the electron (hole) may be used
for a study of the thermodynamic properties of quantum structures with dot
(antidot) in a magnetic field. The temperature dependence of the energy
levels are calculated using the Table 1 at \ any temperature $T$ $($Kelvin).

\acknowledgments The partial support provided by T\"{U}B\.{I}TAK is highly
acknowledged.\newpage

{\normalsize 
}

\bigskip

\bigskip

\baselineskip= 2\baselineskip
\bigskip \newpage

\bigskip

{\normalsize 
}

\baselineskip= 2\baselineskip

\begin{table}[tbp]
\caption{Calculated $f(T)$ with different values of temperature for GaAs. }%
\begin{tabular}{llll}
\tableline\tableline$T$ $(K)$ & $\mu (T)/\mu _{e}$ & $T$ $(K)$ & $\mu
(T)/\mu _{e}$ \\ 
\tableline$0$ & $0.0669984$ & $170$ & $0.0653679$ \\ 
$10$ & $0.0669886$ & $180$ & $0.0652177$ \\ 
$20$ & $0.0669608$ & $190$ & $0.0650643$ \\ 
$30$ & $0.0669174$ & $200$ & $0.0649080$ \\ 
$40$ & $0.0668603$ & $210$ & $0.0647490$ \\ 
$50$ & $0.0667911$ & $220$ & $0.0645874$ \\ 
$60$ & $0.0667112$ & $230$ & $0.0644235$ \\ 
$70$ & $0.0666217$ & $240$ & $0.0642573$ \\ 
$80$ & $0.0665236$ & $250$ & $0.0640891$\tablenotemark[1] \\ 
$90$ & $0.0664178$ & $260$ & $0.0639188$ \\ 
$100$ & $0.0663051$ & $270$ & $0.0637468$ \\ 
$110$ & $0.0661861$ & $280$ & $0.0635730$ \\ 
$120$ & $0.0660614$ & $290$ & $0.0633976$ \\ 
$130$ & $0.0659315$ & $300$ & $0.0632206$\tablenotemark[1] \\ 
$140$ & $0.0657968$ & $350$ & $0.0623154$ \\ 
$150$ & $0.0656577$\tablenotemark[1]\tablenotetext[1]{See Ref. [28].} & $400$
& $0.0613818$\tablenotemark[1] \\ 
$160$ & $0.0655147$ & $500$ & $0.0594513$\tablenotemark[1] \\ 
\tableline &  &  & 
\end{tabular}%
\end{table}

\FRAME{ftbpFO}{0.0277in}{0.0277in}{0pt}{\Qct{Pseudodot $n=0$ energy levels
(in $\hbar \protect\omega _{c}$ unit) as a function of magnetic quantum
number $m$ in the presence and absence of PHQD potential ($a=12$ and $a=0$)
and in the presence and absence of AB flux field ($\protect\xi =8$ and $%
\protect\xi =0$) for magnetic field $B=6$ $T.$}}{}{Figure 1}{}

\FRAME{ftbpFO}{0.0277in}{0.0277in}{0pt}{\Qct{Ground state pseudodot energy
levels (in $meV$) as a function of magnetic field $B$ (in $Tesla$). Solid,
dotted and dashed curves represent the pseudodot energy levels in presence
of AB flux field, Landau energy levels in presence of AB flux field and
Landau levels in the absence of AB flux field, respectively \ for magnetic
quantum number $m=27$.}}{}{Figure 2}{}\FRAME{ftbpFO}{0.0277in}{0.0277in}{0pt%
}{\Qct{Ground state pseudodot energy levels (in $meV$) as a function of
magnetic field $B$ (in $Tesla$). Solid, dotted and dashed curves represent
the pseudodot energy levels in presence of AB flux field, Landau energy
levels in presence of AB flux field and Landau levels in the absence of AB
flux field, respectively \ for magnetic quantum number $m=35$.}}{}{Figure 3}{%
}\FRAME{ftbpFO}{0.0277in}{0.0277in}{0pt}{\Qct{Ground state pseudodot energy
levels (in $meV$) as a function of magnetic field $B$ (in $Tesla$). Solid,
dotted and dashed curves represent the pseudodot energy levels in presence
of AB flux field, Landau energy levels in presence of AB flux field and
Landau levels in the absence of AB flux field, respectively \ for magnetic
quantum number $m=1$.}}{}{Figure 4}{}\FRAME{ftbpFO}{0.0277in}{0.0277in}{0pt}{%
\Qct{Ground state pseudodot energy levels (in $meV$) as a function of
magnetic field $B$ (in $Tesla$). Solid, dotted and dashed curves represent
the pseudodot energy levels in presence of AB flux field, Landau energy
levels in presence of AB flux field and Landau levels in the absence of AB
flux field, respectively \ for magnetic quantum number $m=0$.}}{}{Figure 5}{}%
\FRAME{ftbpFO}{0.0277in}{0.0277in}{0pt}{\Qct{Ground state pseudodot energy
levels (in $meV$) as a function of magnetic field $B$ (in $Tesla$). Solid,
dotted and dashed curves represent the pseudodot energy levels in presence
of AB flux field, Landau energy levels in presence of AB flux field and
Landau levels in the absence of AB flux field, respectively \ for magnetic
quantum number $m=-24$.}}{}{Figure 6}{}\FRAME{ftbpFO}{0.0277in}{0.0277in}{0pt%
}{\Qct{Ground state pseudodot energy levels (in $meV$) as a function of
magnetic field $B$ (in $Tesla$). Solid, dotted and dashed curves represent
the pseudodot energy levels in presence of AB flux field, Landau energy
levels in presence of AB flux field and Landau levels in the absence of AB
flux field, respectively \ for magnetic quantum number $m=-16$.}}{}{Figure 7%
}{}\bigskip \FRAME{ftbpFO}{0.0277in}{0.0277in}{0pt}{\Qct{Ground state GaAs
pseudodot energy levels (in $meV$) as a function of magnetic field $B$ (in $%
Tesla$). Solid, dotted and dashed curves represent the pseudodot energy
levels in presence of AB flux field, Landau energy levels in presence of AB
flux field and Landau levels in the absence of AB flux field, respectively \
for magnetic quantum number $m=0$.}}{}{Figure 8}{}\FRAME{ftbpFO}{0.0277in}{%
0.0277in}{0pt}{\Qct{GaAs pseudodot energy levels (in $meV$) as a function of
magnetic field $B$ (in $Tesla$). Solid, dotted and dashed curves represent
the pseudodot energy levels in presence of AB flux field, Landau energy
levels in presence of AB flux field and Landau levels in the absence of AB
flux field, respectively \ for $n=5$ and $m=0$.}}{}{Figure 9}{}\FRAME{ftbpFO%
}{0.0277in}{0.0277in}{0pt}{\Qct{Ground state GaAs pseudodot energy levels
(in $meV$) as a function of magnetic field $B$ (in $Tesla$). Solid, dotted
and dashed curves represent the pseudodot energy levels in presence of AB
flux field, Landau energy levels in presence of AB flux field and Landau
levels in the absence of AB flux field, respectively \ for magnetic quantum
number $m=5$.}}{}{Figure 10}{}\FRAME{ftbpFO}{0.0277in}{0.0277in}{0pt}{\Qct{%
Ground state GaAs pseudodot energy levels (in $meV$) as a function of
magnetic field $B$ (in $Tesla$). Solid, dotted and dashed curves represent
the pseudodot energy levels in presence of AB flux field, Landau energy
levels in presence of AB flux field and Landau levels in the absence of AB
flux field, respectively \ for magnetic quantum number $m=-5$.}}{}{Figure 11%
}{}\FRAME{ftbpFO}{0.0277in}{0.0277in}{0pt}{\Qct{GaAs pseudodot energy levels
(in $meV$) as a function of magnetic field $B$ (in $Tesla$). Solid, dotted
and dashed curves represent the pseudodot energy levels in presence of AB
flux field, Landau energy levels in presence of AB flux field and Landau
levels in the absence of AB flux field, respectively \ for $n=5$ and $m=-5$.}%
}{}{Figure 12}{}

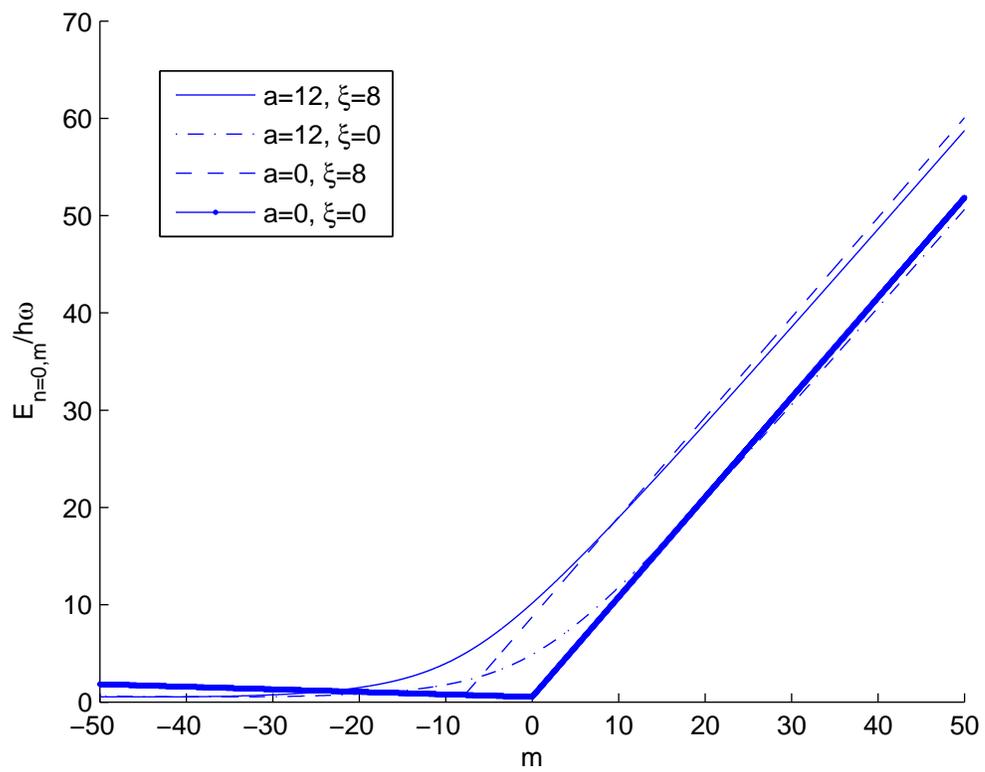
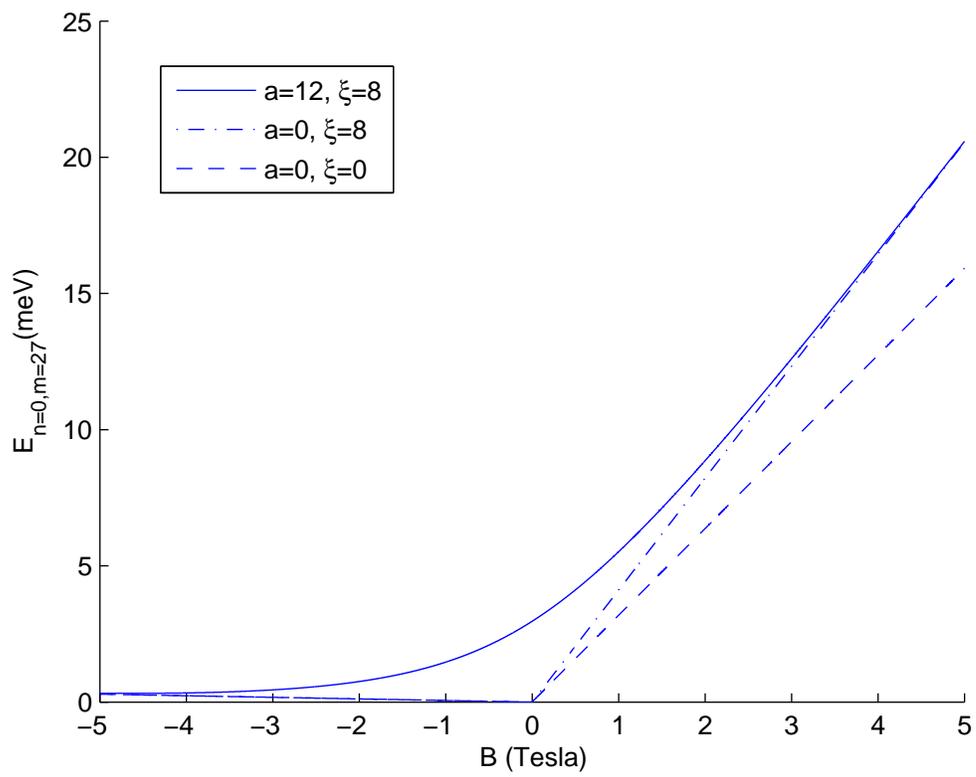
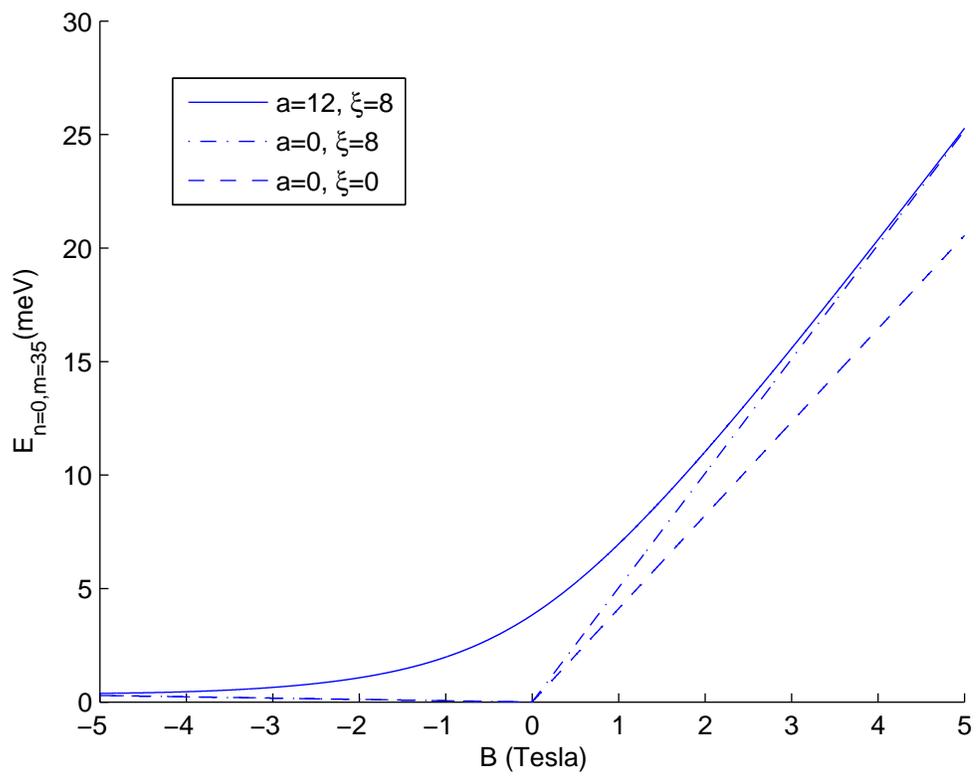
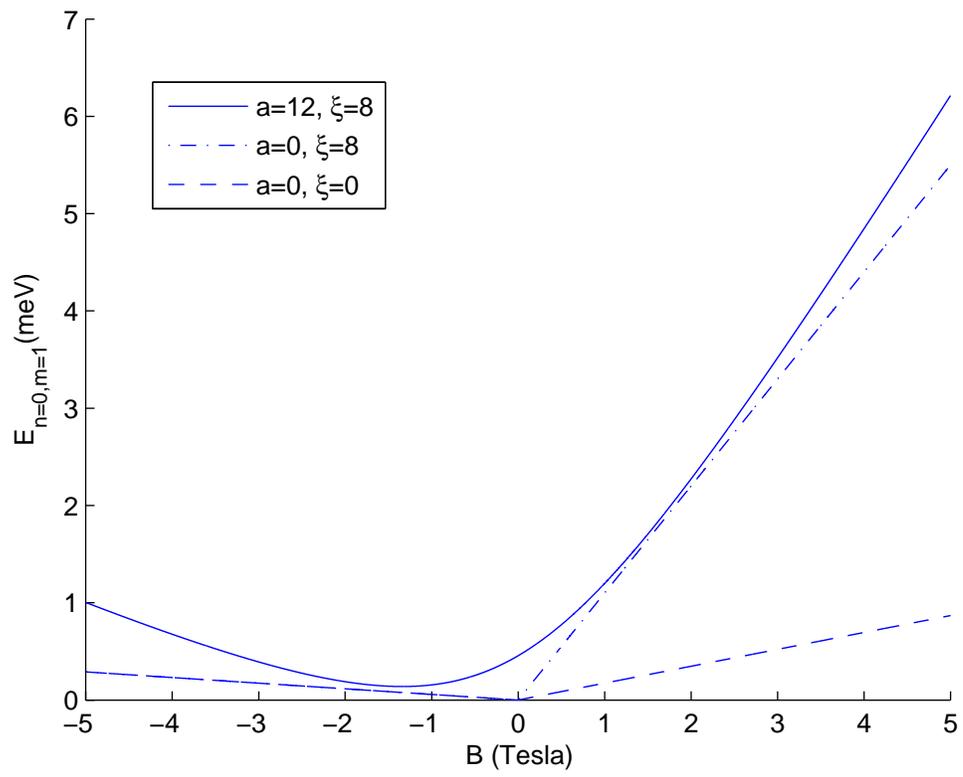
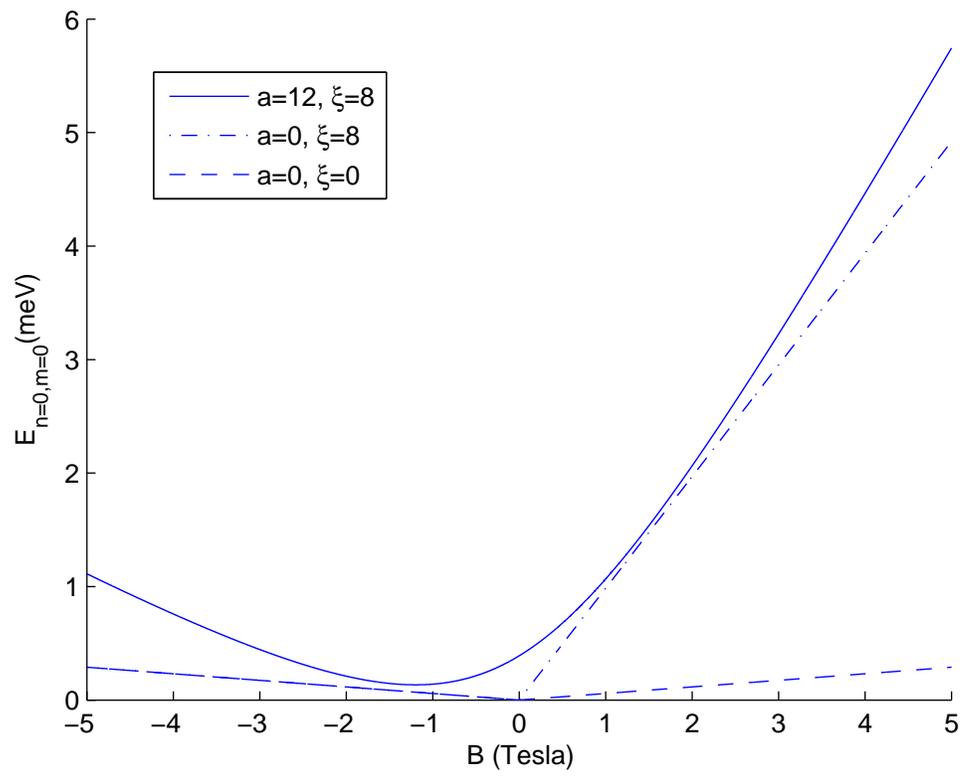
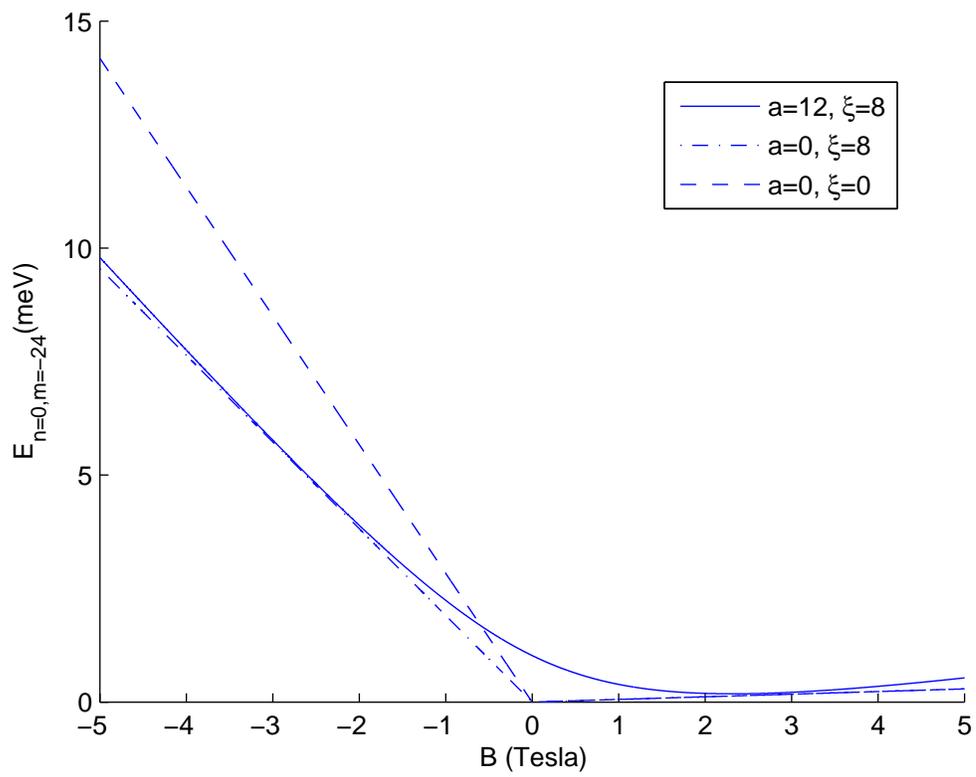
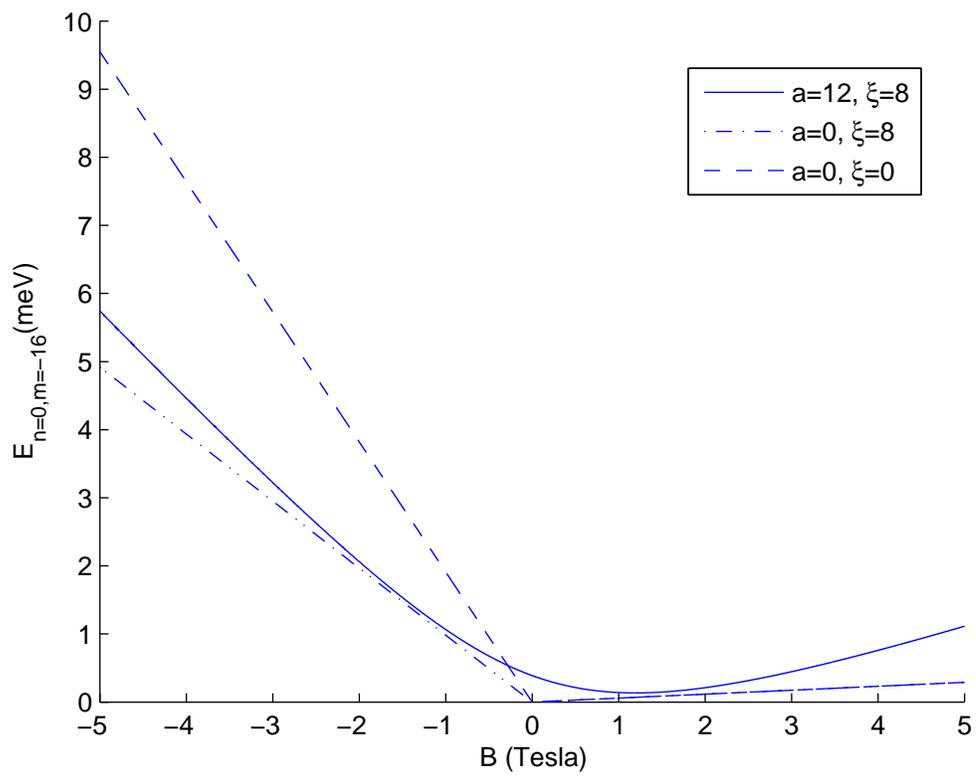
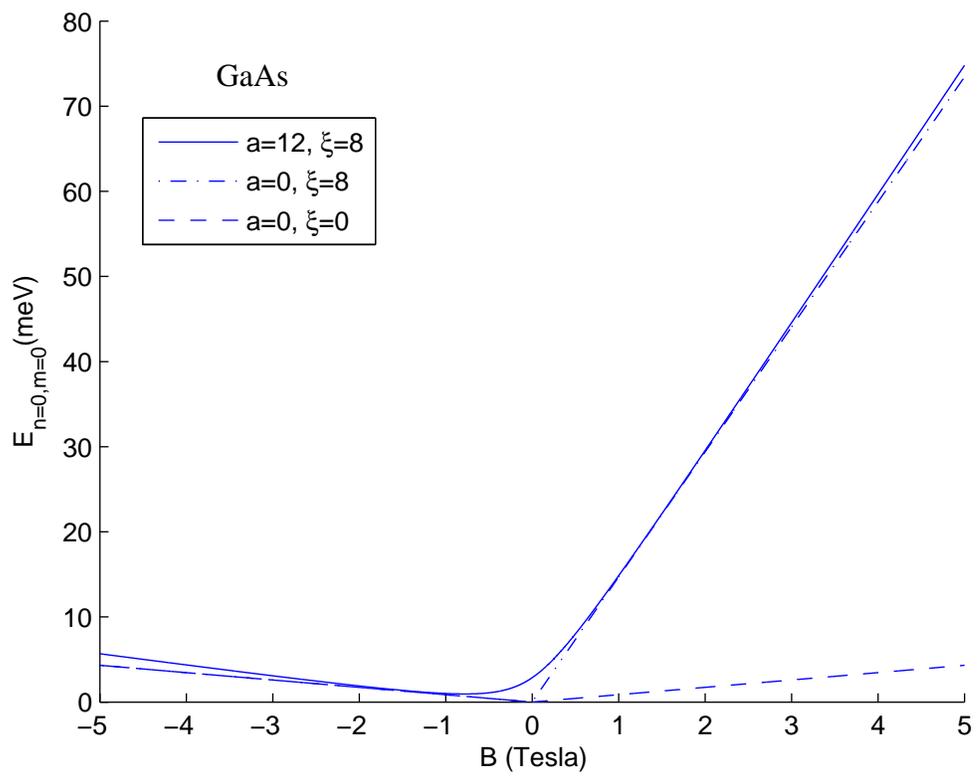
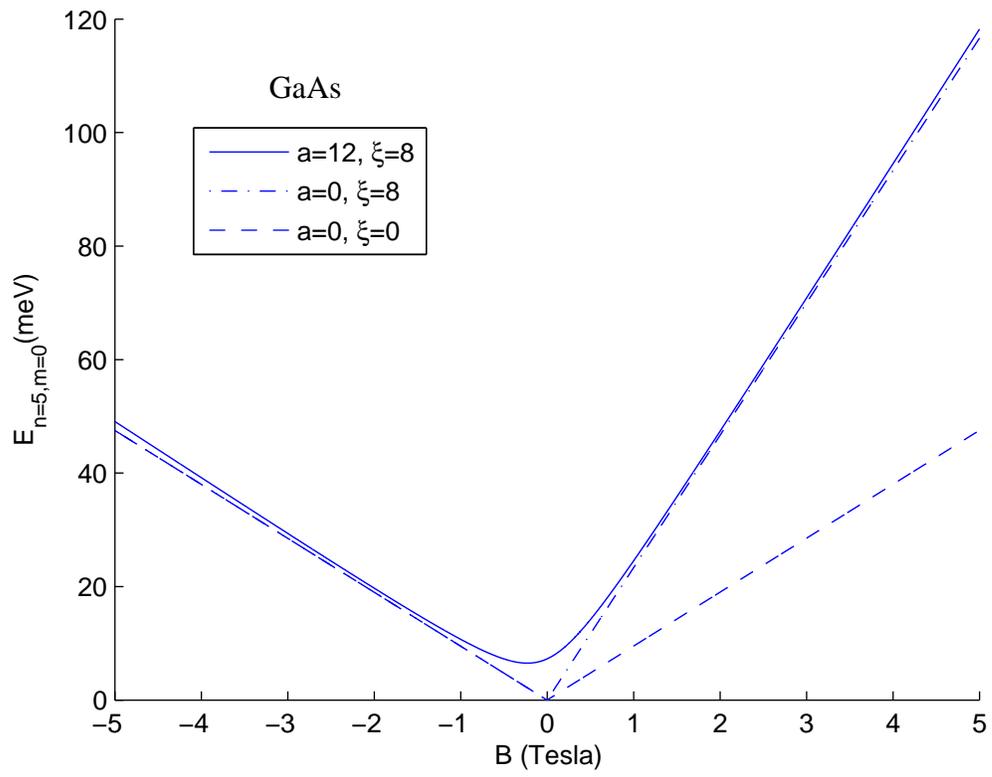
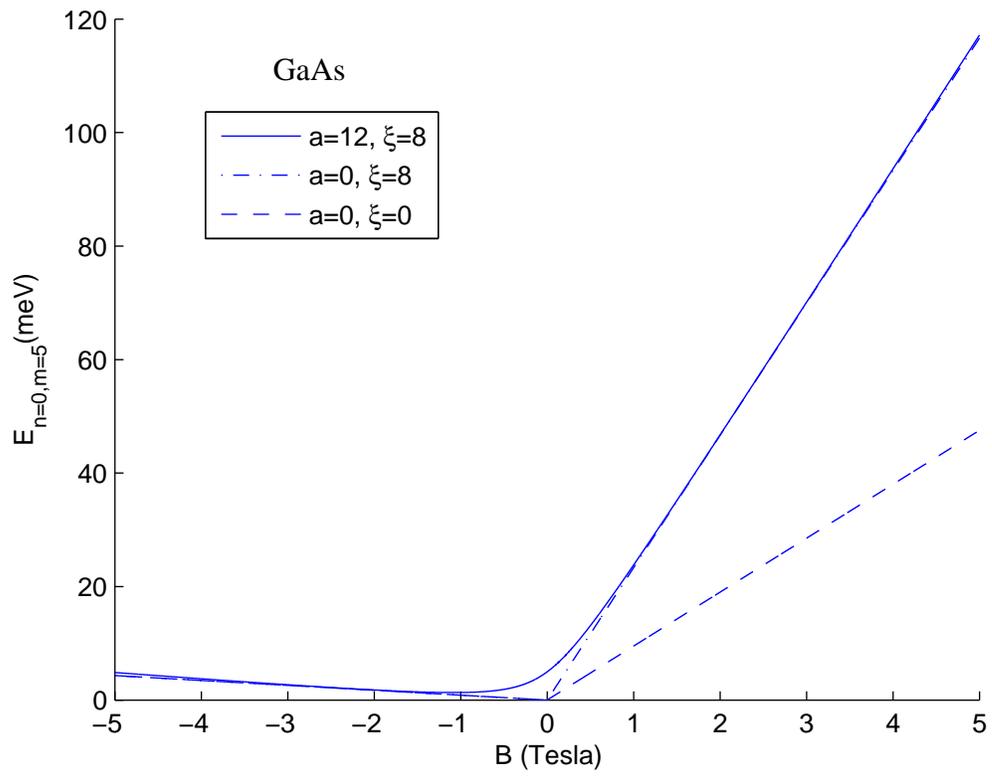
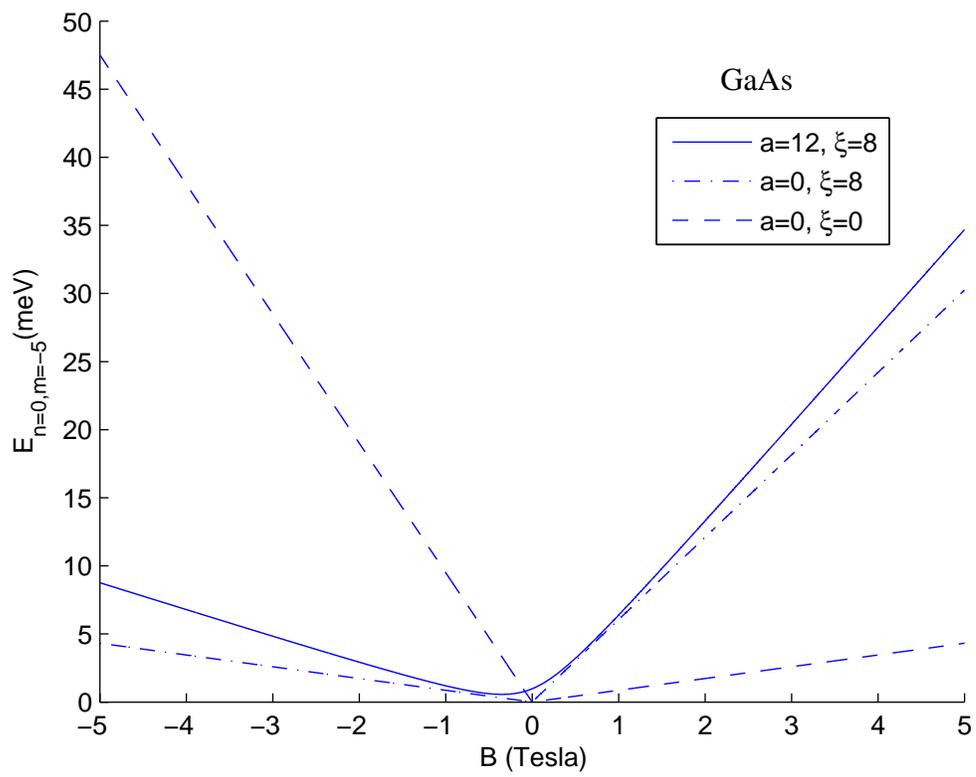
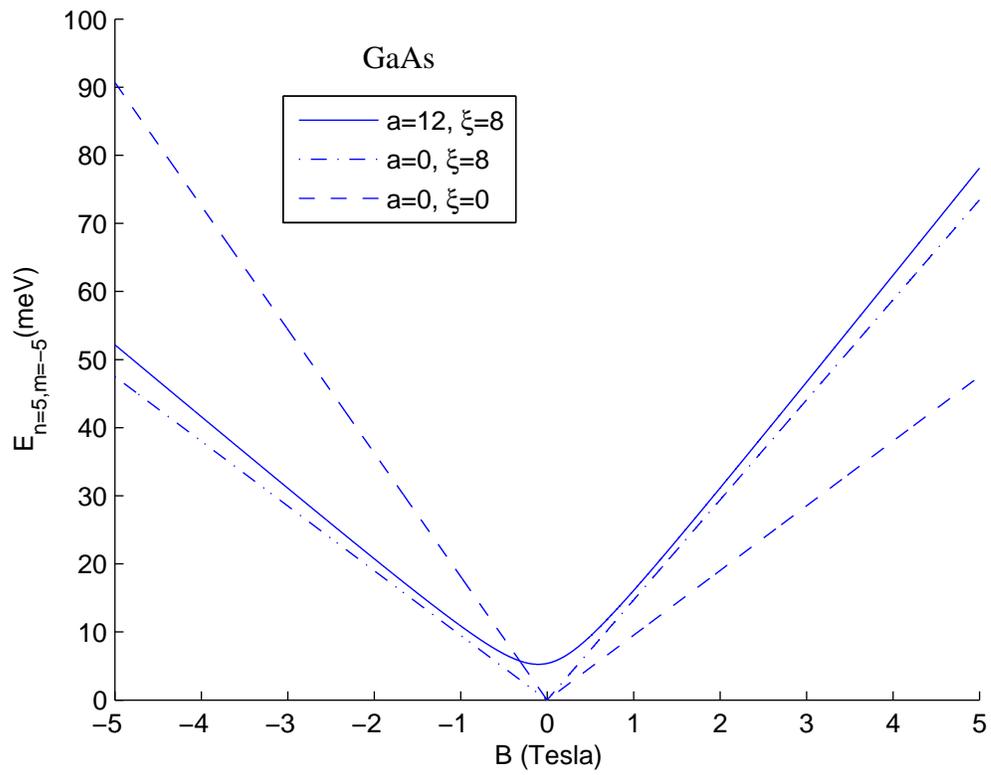
\end{document}